\documentclass{pasj00}
\begin{document}
\SetRunningHead{N. Ota et al.}{Suzaku Observations of the Centaurus Cluster} 
\Received{}%{yyyy/mm/dd}
\Accepted{}%{yyyy/mm/dd}

\title{Suzaku Observations of the Centaurus Cluster: 
Absence of Bulk Motions in the Intracluster Medium}

\author{%
Naomi \textsc{Ota},\altaffilmark{1}
Yasushi \textsc{Fukazawa}, \altaffilmark{2}
Andrew C. \textsc{Fabian}, \altaffilmark{3}
Takehiro \textsc{Kanemaru}, \altaffilmark{4,1} \\
Madoka \textsc{Kawaharada}, \altaffilmark{5}
Naomi \textsc{Kawano}, \altaffilmark{2}
Richard L. \textsc{Kelley}, \altaffilmark{6}
Takao \textsc{Kitaguchi}, \altaffilmark{5}\\
Kazuo \textsc{Makishima},\altaffilmark{5,1} 
Kyoko \textsc{Matsushita}, \altaffilmark{4}
Kouichi \textsc{Murase},\altaffilmark{7}
Kazuhiro \textsc{Nakazawa},\altaffilmark{8}\\
Takaya \textsc{Ohashi},\altaffilmark{9} 
Jeremy S. \textsc{Sanders},\altaffilmark{3}
Takayuki \textsc{Tamura},\altaffilmark{8} 
and 
Yuji \textsc{Urata} \altaffilmark{7}
}

\altaffiltext{1}{Cosmic Radiation Laboratory, 
 RIKEN, 2-1 Hirosawa, Wako, Saitama 351-0198}
\email{ota@crab.riken.jp}
\altaffiltext{2}{Department of Physical Science, 
   School of Science, Hiroshima University, \\
   1-3-1 Kagamiyama, Higashi-Hiroshima, Hiroshima 739-8526}
\altaffiltext{3}{Institute of Astronomy, Madingley Road, Cambridge CB3 0HA, UK}
\altaffiltext{4}{Department of Physics, Tokyo University of Science, 
   1-3 Kagurazaka, Shinjuku-ku, Tokyo 162-8601}
\altaffiltext{5}{Department of Physics, University of Tokyo, 
   7-3-1 Hongo, Bunkyo-ku, Tokyo 113-0033}
\altaffiltext{6}{NASA/Goddard Space Flight Center, Code 662, Greenbelt, MD 20771, USA}
\altaffiltext{7}{Department of Physics, Saitama University, 
   Shimo-Okubo, Saitama 338-8570}
\altaffiltext{8}{Institute of Space and Astronautical Science (ISAS/JAXA),\\
   3-1-1 Yoshinodai, Sagamihara, Kanagawa 229-8510}
\altaffiltext{9}{Department of Physics, Tokyo Metropolitan University, 
   1-1 Minami-Osawa, Hachioji, Tokyo 192-0397}
 

\maketitle

\begin{abstract}
The Centaurus cluster ($z=0.0104$) was observed with the X-ray Imaging
Spectrometer (XIS) onboard the Suzaku X-ray satellite in three
pointings, one centered on the cluster core and the other two offset
by $\pm 8\arcmin$ in declination. To search for possible bulk motions
of the intracluster medium, the central energy of He-like Fe-K line
(at a rest-frame energy of 6.7~keV) was examined to look for a
positional dependence. Over spatial scales of 50~kpc to 140~kpc around
the cluster core, the central line energy was found to be constant
within the calibration error of 15~eV.  The 90\% upper limit on the
line-of-sight velocity difference is $|\Delta v|<1400~{\rm
  km\,s^{-1}}$, giving a tighter constraint than previous
measurements.  The significant velocity gradients inferred from a
previous Chandra study were not detected between two pairs of
rectangular regions near the cluster core. These results suggest that
the bulk velocity does not largely exceed the thermal velocity of the
gas in the central region of the Centaurus cluster.  The mean redshift
of the intracluster medium was determined to be $0.0097$, in agreement
with the optical redshift of the cluster within the calibration
uncertainty.  Implications of the present results for the estimation
of the cluster mass are briefly discussed.
\end{abstract}

\section{Introduction}

Clusters of galaxies are the most massive  gravitationally bound
structures in the Universe, and are thought to grow into larger
systems through complex interactions between smaller systems.
Signatures of such merging events may manifest themselves in
non-Gaussian velocity distributions of member galaxies, temperature
and density inhomogeneities in the intracluster medium (ICM), and bulk
motions of the ICM. Indeed, numerical simulations predict the
existence of bulk ICM flows with a substantial fraction of the virial
velocity ($\sim 1000~{\rm km\,s^{-1}}$ for rich clusters), lasting
several Gyr after each merger event (e.g., \cite{Roettinger_etal_1996,
  Norman_Bryan_1999}).   
Therefore, measurements of the ICM velocity structure
provide very important information 
with which to understand the cluster dynamics. 

 If the ICM has a significant bulk velocity compared to its thermal
 velocity, the associated non-thermal pressure would endanger the
 assumption of hydrostatic ICM equilibrium, which is usually assumed
 in deriving the total gravitating mass in a cluster.  For example,
 the factor 2--3 discrepancy between the X-ray and lensing mass
 estimations in some objects (e.g., \cite{Ota_etal_2004,
   Hattori_etal_1999}) could be due to this effect.  Therefore, the
 presence/absence of ICM bulk motion has a great impact on the cluster
 studies, including their mass estimates and cosmological
 applications.

We are thus encouraged to observationally constrain the ICM motion.
Such an attempt may be carried out by examining X-ray emission lines
from heavy elements for Doppler shifts and broadenings (e.g.,
\cite{Dupke_Bregman_2001}), or by utilizing kinetic Sunyaev-Zel'dovich
effects in the radio wavebands (e.g., \cite{Sunyaev_etal_2003}). For
the former method, X-ray spectroscopy of the emission lines from heavy
ions, particularly the $6.7$~keV K-lines from Helium-like iron atoms,
provides the most sensitive way of studying such ICM motions.  By
observing X-ray spectra in the 0.5--8.5 keV range from the core
regions of the Centaurus cluster (Abell~3526) with the Chandra ACIS,
\citet{Dupke_Bregman_2006} recently reported on a large velocity
gradient of $2400\pm1000~{\rm km\,s^{-1}}$ over a spatial scale of
100~kpc, reconfirming their previous ASCA measurements
\citep{Dupke_Bregman_2001}.  However, their results might still be
subject to residual uncertainties arising from temporal and intrachip
gain variability, requiring an independent cross confirmation.  Since
a bulk velocity of $1000~{\rm km\,s^{-1}}$ along the line of sight
translates to a shift in the He-like Fe-K line energy of $\Delta E =
22$~eV, we need not only sufficient photon statistics and good energy
resolution, but also precise instrumental gain calibrations with an
accuracy better than $\sim 20$ eV (or $\sim 0.3\%$).

The Centaurus cluster (Abell~3526), at the redshift of $z_{\rm
  cl}=0.0104$, is a nearby X-ray bright cluster of galaxies, hosting
the cD galaxy NGC~4696.  It has been extensively studied at various
wavelengths.  \citet{Lucey_etal_1986} found a bimodal velocity
distribution of galaxies, corresponding to the main cluster (Cen~30)
centered on NGC~4696, and a subcluster (Cen~45) centered on NGC~4709,
with a velocity difference of $\sim 1500~{\rm km\,s^{-1}}$.  They
hence argued that Cen~30 is currently accreting Cen~45.  The X-ray
measured ICM temperature is $\sim 3.8$~keV in outer regions, and
decreases toward the center to $\sim 1.7$~keV (e.g.,
\cite{Ikebe_etal_1999, Makishima_etal_2001, Takahashi_etal_2004}).
Although the central part of this cluster is relatively relaxed (e.g.,
\cite{Allen_Fabian_1994}), at larger radii the temperature maps
derived with ASCA reveal a hot spot coincident with Cen~45
\citep{Churazov_etal_1999, Furusho_etal_2001}, supporting the idea
that there is a past or on-going merger episode.  The central region
is very rich in metals, with a large abundance gradient
\citep{Fukazawa_etal_1994, Ikebe_etal_1999, Graham_etal_2006,
  Sanders_Fabian_2006}.  Emitting very intense Fe-K emission lines,
the central region of the Centaurus cluster provides an excellent
opportunity for a study of ICM bulk-motion.

In this paper, we present an analysis of the bulk velocity of the gas
over the central 240 kpc of the Centaurus cluster, conducted with the
X-ray Imaging Spectrometer (XIS; \cite{Koyama_etal_2006}) on board the
Suzaku satellite \citep{Mitsuda_etal_2006}.  The XIS has an excellent
sensitivity at the iron K-line energies, and a very low background.
In addition, the current accuracy of the XIS in-orbit calibration is
already high enough to enable us to make meaningful searches for
possible ICM bulk motions.  The Suzaku X-ray telescope (XRT;
\cite{Serlemitsos_etal_2006}) has a spatial resolution of $\sim
2\arcmin$ (Half Power Diameter; HPD), allowing us to study spatial
velocity variations on reasonable scales ($\sim$a few arcminutes). We
therefore test the previously claimed large velocity gradient near the
cluster center.

Throughout this paper we adopt $\Omega_{\rm M}=0.3$,
$\Omega_{\Lambda}=0.7$, and $H_0=70~{\rm km\, s^{-1}\,Mpc^{-1}}$;
$1\arcmin$ corresponds to 13.5~kpc at the cluster redshift of $z_{\rm
  cl}=0.0104$.  The quoted statistical errors refer to 68\% confidence
ranges, unless otherwise stated.

\section{Observation}

\begin{figure}
  \begin{center}
    \FigureFile(90mm,90mm){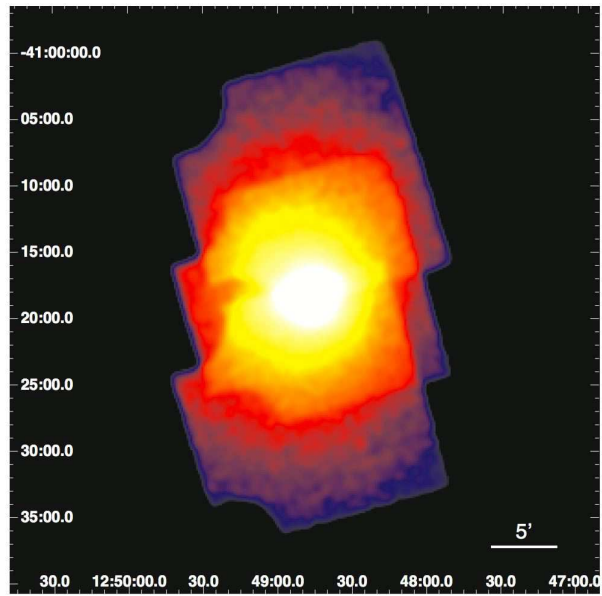}
  \end{center}
  \caption{Suzaku XIS0 smoothed image of the Centaurus cluster in the
    0.4--10~keV band.  The three pointing images are 
      superimposed, but are not corrected for the overlapping
    exposure.  Two corners of the CCD chip illuminated by the onboard
    calibration sources are excluded.  The X-ray emission is peaked at
    NGC~4696.  The subcluster centered on NGC~4709 is about
    $15\arcmin$ away from the center and outside the XIS fields of
    view.}
    \label{fig1}
\end{figure}

Three pointed observations of the Centaurus cluster were conducted
between 2005 December 27--29, in the Suzaku Phase-I period: one
centered on the cD galaxy, NGC~4696, and the other two (Offset1 and
Offset2) offset by $- 8\arcmin$ and $+8\arcmin$ in declination,
respectively.  The details of the observations is given in
table~\ref{tab1}.  The two onboard instruments, the XIS and the Hard
X-ray Detector (HXD; \cite{Takahashi_etal_2006}), were operated in
their normal modes during the observations.  The XIS consists of four
X-ray sensitive CCD cameras: three front-illuminated (FI) CCDs and one
back-illuminated (BI) CCD.  In the present paper, we use data from the
three FI-chip cameras (XIS0, 2, and 3), since they have higher
sensitivities as well as lower backgrounds at the iron-K line
energies, compared to the BI-chip camera (XIS1).  As shown in
figure~\ref{fig1}, the extended ICM emission was clearly detected, the
center of which $(\alpha,\delta)=(192^\circ.2058, -41^\circ.30424)$ in
J2000 coordinates, is consistent within $\sim 25\arcsec$ of the
position of NGC~4696.

%%%%%%%%%%%%%%%%%%%%%%%%%%%%%%%%%%%%%%%%%%%%
\begin{table*}
\begin{center}
\caption{Details of the Suzaku observations of the Centaurus cluster}
\label{tab1}
\begin{tabular}{lllllll}\hline\hline
Target name  & Sequence No. & Date 
& \multicolumn{2}{c}{Coordinates\footnotemark[*]} & Roll angle 
& Exposure\footnotemark[$\dagger$] \\
& & & $\alpha$ (deg) & $\delta$ (deg) & (deg) & (s) \\ \hline
CENTAURUS\_CLUSTER & 800014010 & 2005 Dec 27 
 & 192.2054  & $-41.3111$ & $-253.76$ & 29397 \\
CENCL\_Offset1 & 800015010 & 2005 Dec 28  
 & 192.2054 & $-41.444$ & $-253.72$ & 33431 \\
CENCL\_Offset2 & 800016010 & 2005 Dec 29  
 & 192.2054 & $-41.178$ & $-253.48$ & 31647 \\ \hline
\end{tabular}
\end{center}
\footnotemark[*] Pointing coordinates in J2000.\\
\footnotemark[$\dagger$] Net exposure time of the XIS sensors after data filtering. 
\end{table*}
%%%%%%%%%%%%%%%%%%%%%%%%%%%%%%%%%%%%%%%%%%%%

%%%%%%%%%%%%%%%%%%%%%%%%%%%%%%%%%%%%%%%%%%%%
%%%%%%%%%%%%%%%%%%%%%%%%%%%%%%%%%%%%%%%%%%%%

The data reduction was performed using HEASOFT version 6.0.6.  The XIS
event lists created in the rev0.7 pipeline processing were filtered
with the following criteria: the geomagnetic cut-off rigidity $>6$~GV,
the elevation angle from the earth limb $>10^{\circ}$, and the
satellite outside the South Atlantic Anomaly.  The net exposures are
29.4, 33.4, and 31.6~ks for the Center, Offset1, and Offset2
pointings, respectively.  We selected XIS events with grades 0, 2, 3,
4, and 6.  The XIS response matrix files,
\verb|ae_xi[0,2,3]_20060213.rmf|, and the on-axis auxiliary files,
\verb|ae_xi[0,2,3]_xisnorm6_20060415.arf| were used in the spectral
analysis.

\section{Data analysis and Results}

Our primary goal is to constrain the line-of-sight bulk velocity of
the ICM using the iron-K line at 6.7~keV.  For this purpose, the
accuracy of photon energy measurements is crucial.  We briefly review
in \S\ref{subsec:accuracy} the current XIS calibration uncertainty,
and describe in
\S\ref{subsec:analysis_small}--\S\ref{subsec:analysis_large} our
analyses of the velocity structure of the ICM on small and large
spatial scales.

\subsection{Accuracy of the energy scale}\label{subsec:accuracy}

In order to examine the accuracy of the XIS energy scale, we checked
the following three key points: the line centroid energies of
calibration sources illuminating some corners of the CCD chips, the
positional gain variations due to Charge Transfer Inefficiency, and
the agreement of line centroid energies among the three pointings onto
the same sky regions of the Centaurus cluster.

The fiducial absolute energy scale of the XIS is provided by
Mn-K$\alpha$ lines from the built-in calibration sources ($^{55}$Fe),
which illuminates two out of four corners of each XIS chip.  For the
three pointings, we fitted these lines with a Gaussian, and confirmed
that the obtained line centroids of XIS0, XIS2, and XIS3 all agree
within 0.02--0.1\% of the expected value of 5.8951~keV.

Although the calibration isotope fixes the fiducial energy scale, the
gain could vary from place to place on the same CCD chip (intrachip
gain variation), due to Charge Transfer Inefficiency (CTI).  The CTI
increases with time due to irradiation by charged particles, enhancing
the intrachip gain variations particularly along the direction of
charge transfer (so called ACT-Y axis).  The CTI characteristics of
the XIS were calibrated in orbit using two line-rich extended sources,
the Cygnus Loop and Sgr C, observed on 2005 November 23--24 and 2006
February 20--23, respectively.  As a result, the gains of the four XIS
sensors have been equalized over the CCD chips to an accuracy of $\pm
0.2$\% \citep{Koyama_etal_2006}.  These results have been used to
produce the rev0.7 data.  As the three observations of the Centaurus
cluster were carried out on successive days between those of the
Cygnus Loop and Sgr C, our analysis should be little affected by any
temporal gain variation.

We further examined the rev0.7 data of two other extended sources,
Abell~1060 and Abell~426, observed on 2005 November 22 and 2006
February 1, respectively.  The ACT-Y dependence of the iron line
energy was then confirmed to be very small, typically within $\pm
0.2$\%.  From these calibrations, we consider that the intrachip gain
variations of the rev0.7 data from the four XIS sensors are at most
$\pm 0.2$\% ($\pm15$~eV) in the ACT-Y direction.  The gain variation
along the ACT-X axis is also within $\pm 0.2$\%, as confirmed with Sgr
C, Abell~1060, and Abell~426.  Quantitative confidence levels of the
systematic error were estimated based on a detailed analysis of
Abell~426, which has the most intense iron-K lines among the celestial
calibrators utilized here.  Since the four XIS sensors are placed on
the focal plane in different orientations with respect to the
satellite coordinate \citep{Koyama_etal_2006}, the positional gain
variation is effectively studied by comparing the line energies of the
same sky regions (namely different detector regions) on the four
sensors.  We thus divided the XIS field of view of the Abell~426
observation into $8\times8$ cells of size $2\arcmin.1\times
2\arcmin.1$, and estimated the $1\sigma$ systematic error on the iron
line energy by calculating the chi-square value, defined as $\chi^2=
\Sigma_{\rm cell} \Sigma_{i=0}^{3} (E_i-\langle E \rangle)^2/
(\sigma_{i}^2 + \sigma_{\rm sys}^2)$.  Here $E_i$ and $\sigma_{i}$ are
the iron-K line centroid energy and the $1\sigma$ statistical error
obtained with XIS-$i$ in a certain cell, respectively, $\langle E
\rangle=\Sigma_{i=0}^{3} E_i/4$ is the mean iron-K energy, and
$\Sigma_{\rm cell}$ sums up all cells.  If a systematic error of
$\sigma_{\rm sys}=10$~eV (i.e., $\pm 0.13$\%) is tentatively employed,
we obtain $\chi^2/\nu=1.1$ for 144 degrees of freedom. This value of
$\sigma_{\rm sys}$ is hence considered appropriate as a measure of the
residual gain non-uniformity.

Utilizing the overlaps between the three pointings of the Centaurus
cluster (figure~\ref{fig1}), we examined whether the central line
energy of the same sky region turns out to be the same when measured
in the different pointings (and hence by different regions on the CCD
chip).  In detail, we derived the iron-line energy of the central
pointing data of the three FI chips from a $8\arcmin.5\times15\arcmin$
region (roughly coincident with southern half of the XIS field of view
in the central pointing), and compared it with that of the same sky
region observed in Offset1.  The results agree with each other within
$0.1$\%.  Similarly, the Fe-K line energy averaged over a northern
$8\arcmin.5\times15\arcmin$ region showed a good agreement ($0.1$\%)
between the central pointing and Offset2.  If we divide a central
$8\arcmin.5\times8\arcmin.5$ square region centered on the cD galaxy
into $4\times4$ cells of size $2\arcmin.1\times 2\arcmin.1$ each, the
energy scales are again in good agreement between the central pointing
and the two offsets; $\lesssim 0.2$\% in 14 cells, and $\lesssim
0.3$\% in 2 cells (with $1\sigma$ statistical errors being $\sim
20$~eV).

Based on these examinations, we decided to take $\pm 0.13$\% ($\pm
0.2$\%) as the systematic error on the XIS energy scale at the 68\%
(90\%) confidence level.  The 68\% systematic error corresponds to
$\pm10$~eV in terms of the iron-K line energy, or $\pm0.0015$ in
redshift, or $\pm470~{\rm km\,s^{-1}}$ in line-of-sight velocity.

\subsection{Analysis of small-scale velocity structure}
\label{subsec:analysis_small}
\begin{figure}
  \begin{center}
    \FigureFile(90mm,90mm){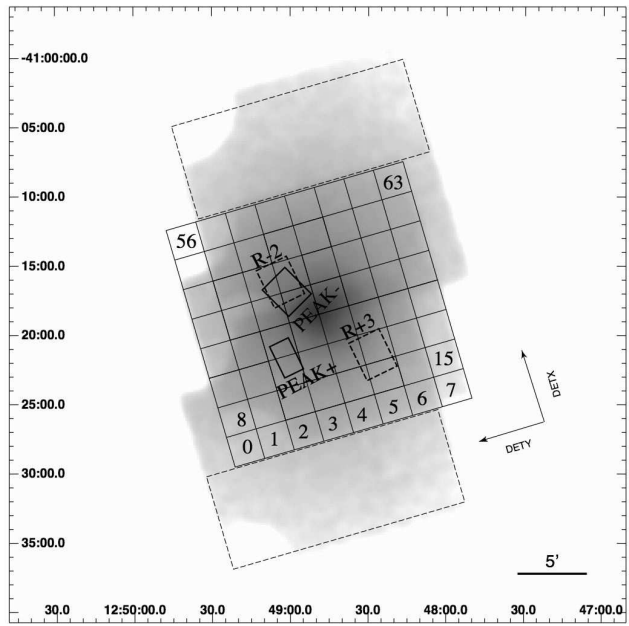}
  \end{center}
  \caption{An image showing the spectral accumulation regions.
    The central $8\times8$ cells, each with a size of $2
    \arcmin.1\times 2\arcmin.1$, are designated as \#0 $\sim$ \# 63,
    and were used for the study of small-scale velocity
    structure.  The two larger rectangular regions (the dashed
    boxes of size $7\arcmin\times18\arcmin$) were used for the
    larger-scale analysis.  The four regions used for comparison
    with the Chandra results \citep{Dupke_Bregman_2006} are also shown
    as PEAK-, PEAK+ (the solid boxes), R-2, and R+3 (the dashed
    boxes).}\label{fig2}
\end{figure}

Now that we have confirmed the XIS energy-scale is accurate, we
proceed to the study of the velocity structure of the ICM in the
Centaurus cluster on a scale of a few arcminutes. For this purpose, we
divided the $18\arcmin\times18\arcmin$ square XIS field of view into
$8\times8$ cells as shown in figure~\ref{fig2}.  Thus, each cell
covers a sky area of $2'.1\times 2'.1$ (=28 kpc $\times$ 28 kpc).
Since the energy scale is consistent among the three XIS sensors and
among the three pointings (\S\ref{subsec:accuracy}), we co-added the
data from XIS0, XIS2, and XIS3, and merged the three pointings where
they overlap.  Excluding the calibration source regions, 52 cells in
total were used in the spectral fitting.  The composite spectra were
rebinned so that each spectral bin had more than 30 counts.

Figure~\ref{fig3} shows an example of the XIS spectrum obtained in
this way, displayed over the iron energy band.  Thus, we observe a
very intense He-like Fe-K$\alpha$ line, a weaker H-like Fe-K$\alpha$
line, and a blend between He-like Ni-K$\alpha$ and He-like Fe-K$\beta$
lines.  In the 5--10~keV band, the source count rate measured in each
cell ranges from $1.6\times10^{-3}$ to $2.5\times10^{-2}$ c~s$^{-1}$
per XIS sensor, with a mean over the 52 cells being
$6.9\times10^{-3}$~c\,s$^{-1}$.  If the energy band is limited to
6.5--6.9~keV, in which the He-like Fe K$\alpha$ line emission is
prominent, these count rates decrease by a factor of 3--5.

%%%%%%%%%%%%%%%%%%%%%%%%%%%%%%%%%%%%%%%%%%%%
%%%%%%%%%%%%%%%%%%%%%%%%%%%%%%%%%%%%%%%%%%%%

Background spectra were derived by integrating events over the same
regions on the detector planes as the on-source data integration,
using a blank-sky observation towards the North Ecliptic Pole
conducted on 2005 September 2--4 for 95~ks.  At the cluster center,
the background surface brightness in the 5--10~keV range is only a few
percent of that of the ICM emission, while the background contribution
increases to $\sim 35$ percent near the edge of the XIS field of view.
The XIS background spectrum has several instrumental fluorescent lines
in the energy band of interest, with the most prominent one being
Ni-K$\alpha$ line at 7.48~keV.  However, as shown in
figure~\ref{fig3}, the background spectrum can be regarded as
featureless in the energy range where the He-like Fe-K$\alpha$ line
from the Centaurus cluster is present.  We subtracted the backgrounds
from the on-source spectra, and confirmed that this process does not
change the Fe-K line energy.
 
\subsection{Modeling the iron-K lines}\label{subsec:specmodel}

To accurately determine the iron-line energies, we fitted the
background-subtracted XIS spectra with a simple model over the 5 -- 10
keV energy range.  The model we chose consists of a continuum
represented by the APEC thermal emission model with metal abundances
reset to zero ($Z=0$), and three Gaussians for the major line
components: the He-like Fe-K$\alpha$ line at 6.70~keV, the H-like
Fe-K$\alpha$ line at 6.97~keV, and a 7.83~keV line representing the
blend of He-like Ni-K$\alpha$ line (7.80~keV) and He-like Fe-K$\beta$
line (7.90~keV).  The quoted line energies refer to their rest-frame
values.

The He-like Fe-K$\alpha$ complex, consisting of 26 lines, exhibits an
asymmetric shape (e.g., \cite{Dubau_Volonte_1980}).  Although the line
profile becomes essentially Gaussian when convolved with the XIS
energy resolution [$\sim 160$~eV (FWHM) at the time of the
  observations of the Centaurus cluster; \cite{Koyama_etal_2006}], the
Gaussian centroid energy, denoted $E_0$, can no longer be identified
with that of the strongest component, namely the resonance line at
6.6986~keV in the rest frame.  To determine the expected value of
$E_0$ when the ICM has no bulk motion and the cluster is at zero
redshift, we produced a simulated spectrum employing the APEC model
and XIS energy response.  Based on ASCA measurements
(\cite{Furusho_etal_2001, Ikebe_etal_1999}), the simulation assumed an
ICM temperature of $kT\sim 3$~keV, and a metallicity of
$Z=1Z_{\odot}$.  By fitting the simulated spectrum with an APEC plus
three Gaussian model as described above, we obtained
$E_0=6.6771\pm0.0001$~keV.  We thus assume the rest-frame He-like
Fe-K$\alpha$ centroid energy to be $E_0=6.677$~keV.

Strictly speaking, the value of $E_0$ is expected to depend upon the
ICM temperature; as the temperature decreases, contributions from
lower-ionization ions increase, making $E_0$ shift towards lower
energies.  For example, the shift is by about 1\% when $kT=1$~keV.
The very center of the Centaurus cluster is known to host a cool
component with a temperature of $kT\sim 1.7$~keV
\citep{Ikebe_etal_1999, Sanders_Fabian_2002, Takahashi_etal_2004}.
However, the effect of the cool plasma on the line-centroid
measurement is estimated to be negligible, because the cool gas is
localized within $1\arcmin$ from the cD galaxy with a relatively minor
emission integral \citep{Ikebe_etal_1999, Sanders_Fabian_2002,
  Takahashi_etal_2004}, and the Fe-K line emissivity of the 1.7~keV
gas is by more than a factor of 2 (7) lower than that of $>2$~keV
($>3$~keV) gas.  Towards higher temperatures, $E_0$ can be regarded as
constant within 0.1\% as long as $2~{\rm keV}<kT<8~{\rm keV}$.

The intrinsic line width of the He-like Fe-K$\alpha$ complex emitted
from the 3~keV gas is derived to be $\sigma_0=27.8\pm 0.2$~eV.  This
is smaller than the instrumental energy resolution at $\sim 7$~keV,
$\sim 70$~eV in Gaussian sigma, and hence a minor contributor to the
observed total line width.  The possibility of the line broadening in
excess of the intrinsic width is discussed in
\S\ref{subsec:turbulent}.

\subsection{Result of spectral fitting}

Using {\tt XSPEC} version~11, we fitted the 5--10~keV spectra from
individual cells with the APEC plus three Gaussian line model
described in \S\ref{subsec:specmodel}.  The three Gaussian energies
were allowed to vary, but their relative energies were constrained to
obey their rest-frame ratios.  Since the XIS energy resolution $\delta
E$ depends on energy $E$ approximately as $\delta E \propto E^{0.5}$
in the energy band of interest, and varies gradually with time
\citep{Koyama_etal_2006}, the Gaussian widths were included as free
parameters, with their relative values constrained to scale with the
square root of the line energies.  The normalizations of the APEC and
three Gaussians were also left free.  Although the analysis utilizes
the $\chi^2$ fitting method, consistent results are derived if
unbinned spectra are fitted by the maximum likelihood method.

\begin{figure}
  \begin{center}
    \FigureFile(80mm,80mm){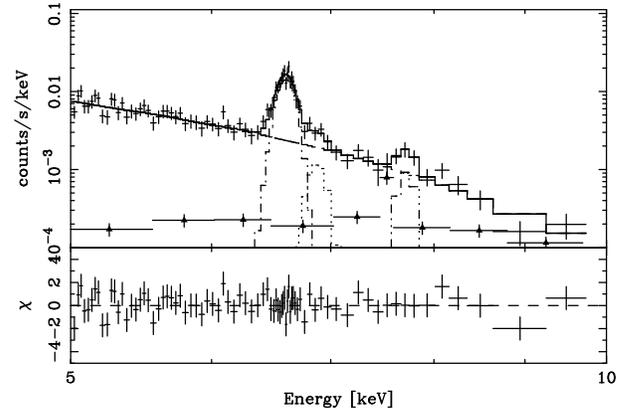}
  \end{center}
  \caption{Background-subtracted XIS0+XIS2+XIS3 spectrum of cell \#44
    in the 5--10~keV band (the crosses), fitted with the APEC model
    with $Z=0$ for the continuum (the dashed line) and three gaussian
    lines; the dash-dot line for the He-like K$\alpha$, the dotted
    line for H-like K$\alpha$, and the dash-dot-dot line for a blend
    of He-like Ni K$\alpha$ and He-like Fe K$\beta$. The best-fit
    total model is shown with the solid line.  The background spectrum
    taken from the North Ecliptic Pole observation is also
      shown (the crosses with triangle).  The bottom panel shows the
    residuals of the fit in terms of the ratio of the residual to the
    uncertainty of each measurement, which is also the contribution to
    $\chi^2$ when squared. } 
  \label{fig3}
\end{figure}

An example of the spectral fit is shown in figure~\ref{fig3}.  We
obtained an acceptable fit at the 90\% confidence level from 49 out of
the 52 cells (except \# 2, 22, and 33), and at the 95\% level from all
the cells.  The reduced $\chi^2$ values, $\chi_\nu^2$, range from
  0.68 to 1.53 in the central $6\times6$ cells, where the number of
  degrees of freedom $\nu$ ranges from 15 to 132.  In the outer 19
  cells, $\chi_\nu^2$ is between 0.30 and and 1.85, and $\nu$ between
  3 and 27. The value of $\nu$ varies as the number of spectral bins
changes, due to the spectral binning criterion.

The line centroids have been constrained in most cases, with the
$1\sigma$ statistical errors of $\sigma_{\rm fit}=3-4$~eV for the
inner $2\times2$ cells, $\sim 4-8$~eV for the surrounding 12 cells,
and $\sim 10$~eV for outer 17 cells surrounding the central $4\times4$
cells.  In most of these 33 cells, the fitted Gaussian widths are
consistent, within the 90\% statistical errors, with those found with
the Mn-K$\alpha$ calibration line ($\sigma_{\rm inst} \sim 40$~eV).
On the other hand, those cells lying near the edge of the field of
view have too poor photon statistics to constrain the line width.  In
analyzing the outermost 19 cells, we thus fixed the line width at
40~eV and the ICM temperature at 3.0~keV.

\begin{figure*}
  \begin{center}
    \FigureFile(160mm,160mm){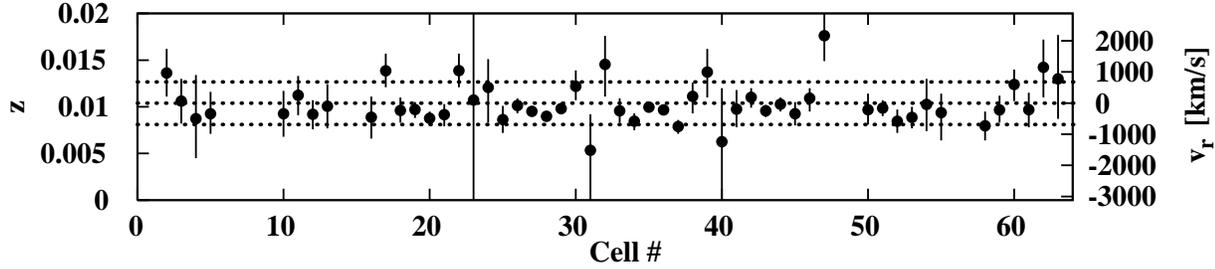}
  \end{center}
  \caption{Results of the redshift measurements using the He-like Fe-K
    lines in 52 cells.  The left and right axes show the
    redshift and radial velocity of the ICM, respectively.  See
    figure~\ref{fig2} for the definitions of the spectral regions.
    The error bars are $1\sigma$ statistical. The optical redshift of
    $z_{\rm cl}=0.0104$ and the range of the present calibrational
    error at the 90\% confidence level, $\pm 0.002$ in redshift (or
    $\pm 700~{\rm km\,s^{-1}}$ in the line-of-sight velocity), are
    indicated with the horizontal dotted lines.  }\label{fig4}
\end{figure*}

The fitted line centroids, $E_{\rm obs}$, are then converted to
redshifts through $z=(E_0 - E_{\rm obs})/E_{\rm obs}$ using
$E_0=6.677$~keV.  The obtained results are plotted in
figure~\ref{fig4}, where the implied ICM radial velocity, $v_r \equiv
c (z - z_{\rm cl})$, is also indicated on the right-side ordinate of
the panels.  As seen from this plot, the measured redshifts are
consistent, in almost all the 52 cells, with the optical value within
the current 90\% calibration errors ($\pm 0.002$ in terms of
redshift).

%%%%%%%%%%%%%%%%%%%%%%%%%%%%%%%%%%%%%%%%%%%%
%%%%%%%%%%%%%%%%%%%%%%%%%%%%%%%%%%%%%%%%%%%%

%%%%%%%%%%%%%%%%%%%%%%%%%%%%%%%%%%%%%%%%%%%%
%%%%%%%%%%%%%%%%%%%%%%%%%%%%%%%%%%%%%%%%%%%%

The weighted mean (and the standard deviation) of the line centroid
measured in the 52 cells is $\langle E_{\rm obs}\rangle=6.6133$~keV
($\pm0.0162$~keV).  The reduced chi-square around this mean is
$\chi^2_{\nu} = \nu^{-1}\Sigma ((E_{\rm obs}- \langle E_{\rm
  obs}\rangle)/\sigma_{\rm fit})^2 =1.1$ with $\nu=51$.  Thus the
measured $E_{\rm obs}$ values are consistent with being constant at
the 90\% confidence level.  The mean redshift (and the standard
deviation) of the ICM is determined to be $\langle z_{\rm obs}\rangle
= 0.0097$ ($\pm 0.0024$), which agrees with the optical redshift of
the cluster $z_{\rm cl}=0.0104$ within the calibrational uncertainty.
For reference, the optical redshift of the cD galaxy is $z_{\rm
  cD}=0.009867$ (the NED database).  The difference between $z_{\rm
  cD}$ and $z_{\rm cl}$, 0.0005, however, is not distinguishable given
the present accuracy of the energy scale.

\subsection{Other systematic errors}\label{subsec:syserr}

The Point Spread Function (PSF) of the Suzaku X-ray telescopes has a
width of $\sim 2\arcmin$ (in half-power diameter), which is comparable
to the size of the spectral regions examined here.  This means that
photons we detect in a certain spectral accumulation region are the
sum of those coming from the corresponding sky region, $C_1$, and
those from its surroundings, $C_2$.  We estimated this effect by the
{\tt xissim} ray-tracing simulator \citep{Ishisaki_etal_2006},
assuming a $kT=3$~keV APEC model for the spectral distribution and a
double $\beta$ model for the surface brightness distribution.
According to \citet{Mohr_etal_1999}, the two $\beta$ components are
assumed to have core radii of $r_{c1}=99~{\rm kpc}$ and
$r_{c2}=8.6~{\rm kpc}$, the same slope parameter $\beta=0.569$, and a
surface brightness ratio of $I_2/I_1=43$ at the center.  This
simulation yielded $C_1 : C_2 \sim 0.4 : 0.6$ in cells lying at
$r<4\arcmin.5$ from the center, and $C_1 : C_2 \sim 0.5 : 0.5$ in
cells surrounding them.  Therefore, the results in figure~\ref{fig4}
remains valid, as long as the spatial resolution of the analysis is
regarded as $\sim 4\arcmin$ ($\sim 50~{\rm kpc}$) instead of $\sim
2\arcmin$.

As reported in \citet{Serlemitsos_etal_2006}, the attitude of
  Suzaku sometimes drifts in both DETX/Y directions with a typical
  peak-to-peak amplitude about $0\arcmin.5-1\arcmin$, mainly
  synchronized with the spacecraft orbital period (96 min).  This
effect blurs the spatial photon distribution on the detector plane,
because the size of the spectral accumulation regions is relatively
small ($\sim 2\arcmin$) in our analysis.  To examine this issue, we
performed Lorentzian fitting to central X-ray surface brightness
profiles projected onto the DETX or DETY axis, and determined the
X-ray centroid position of the cluster every 256 seconds, without
relying on the attitude solution.  As a result, the amplitudes of the
drift were found to be small in this particular observation:
$\Delta{\rm DETX}\sim 37\arcsec$ and $\Delta{\rm DETY}\sim 43\arcsec$.
We produced a mock attitude file (that describes temporal change of
the satellite's euler angles) assuming that the cluster centroid
position moves sinusoidally with a peak-to-peak amplitude of
$40\arcsec$ in the DETX/Y directions and the 96-min period.  By
repeating the same ray-tracing simulation incorporating this mock
attitude file, the effect of the attitude drift on the photon counts
($C_1$ or $C_2$) was estimated to be $\lesssim10$\%, and typically
$C_1 : C_2 \sim 0.4 : 0.6$ inside the XIS field of view.  Therefore
the effect of attitude drift is negligible compared to the original
PSF effects.

\subsection{Analysis of larger-scale velocity structure}
\label{subsec:analysis_large}
In order to search the X-ray data for larger-scale velocity gradients
in the ICM, we compared the redshift in the following three regions;
the central $18\arcmin\times18\arcmin$ square, and two $7\arcmin
\times 18\arcmin$ rectangular regions which are offset
northwards/southwards by about $12\arcmin$ (two dashed boxes in
figure~\ref{fig2}).  The spectral fits in the 5--10~keV band yield
$z=0.00993\pm0.00015 \,(\pm 0.0015)$, $0.0119\pm0.0011\,(\pm 0.0015)$,
and $0.0094\pm0.0015\,(\pm 0.0015)$, for the central, northern offset,
and southern offset regions, respectively.  The 68\% statistical
errors and (the 68\% systematic errors) are quoted.  The radial
velocity is then $v_r=-144\pm 44\,(\pm 470)~{\rm km\,s^{-1}}$,
$441\pm316\,(\pm 470)~{\rm km\,s^{-1}}$, and $-306\pm450\,(\pm
470)~{\rm km\,s^{-1}}$ for the three regions, respectively. Thus, no
significant velocity variation on $\sim 10$-arcminutes scales is seen
from the above analysis. Taking the sum of the statistical and
systematic errors in quadrature, we estimate the 90\% upper limit on
the velocity gradient among the above three regions to be $|\Delta v|
< 1400~{\rm km\,s^{-1}}$.

\section{Discussion}

 From the careful analysis of the high-quality Fe-K line data
  obtained with the Suzaku XIS, we found no evidence of significant
  spatial gradients in the Fe-K line center energies inside the XIS
  fields of view.  The derived upper limits apparently exclude the
  large velocity difference previously claimed by
  \citet{Dupke_Bregman_2006}.  In \S\ref{subsec:comparison_chandra},
  we further investigate this issue by analyzing the same regions as
  studied by \citet{Dupke_Bregman_2006}.  In
  \S\ref{subsec:implications}, implications of the present results are
  discussed, followed by \S\ref{subsec:turbulent} where we constrain
  Doppler broadening of the Fe-K lines in search for turbulent motions
  of the ICM.

\subsection{Comparisons with previous results}
\label{subsec:comparison_chandra}

A large velocity difference by $\sim 2500~{\rm km\,s^{-1}}$ was
reported in the core of the Centaurus cluster based on the previous
ASCA and Chandra observations \citep{Dupke_Bregman_2001,
  Dupke_Bregman_2005, Dupke_Bregman_2006}.  Specifically, by comparing
the redshifts measured with the Chandra/ACIS-S3 spectra in various
regions near the cluster center, \citet{Dupke_Bregman_2006} obtained
(1) a maximum velocity gradient of $\Delta v=2900\pm700~{\rm
  km\,s^{-1}}$ between a pair of rectangular regions,
$2\arcmin.3\times2\arcmin.7$ (PEAK-) and $1\arcmin.5\times2\arcmin.5$
(PEAK+), and (2) another large difference by $\Delta
v=2400\pm1000~{\rm km\,s^{-1}}$ between two $2\arcmin.4\times
3\arcmin$ regions (R-2 and R+3).  We show these four regions on our
figure~\ref{fig2}.

Since the report by \citet{Dupke_Bregman_2006} apparently disagrees
with our results, we examined the present Suzaku data for the above
two cases, by accumulating spectra from the same sky regions on the
three FI chips.  (As seen in figure~\ref{fig2}, the PEAK- and PEAK+
regions are located near cells \#35 and \#18, and R-2 and R+3 are near
cells \#43 and \#21 in our analysis.)  After co-adding the central and
offset data, the total exposure time is 183~ks for PEAK- and R-2, and
188~ks for PEAK+ and R+3; these are about 5 times longer than the
Chandra exposure.

%%%%%%%%%%%%%%%%%%%%%%%%%%%%%%%%%%%%%%%%%%%%
\begin{table*}
\begin{center}
\caption{Comparisons of velocity differences measured with Suzaku and Chandra}
\label{tab2}
\begin{tabular}{llll}\hline\hline
Region           & Suzaku/XIS & Chandra/ACIS \\ \cline{2-2} \cline{3-3}
   & $\Delta v$~(${\rm km\,s^{-1}}$) \footnotemark[*] 
   & $\Delta v$~(${\rm km\,s^{-1}}$)\footnotemark[$\dagger$] \\ \hline
PEAK-, PEAK+ &  $-660 \pm 390\,(\pm 660)$ & $2900\pm700$\\
R-2, R+3            &  $-540 \pm 360\,(\pm 660)$ & $2400\pm1000$ \\ \hline
\end{tabular}
\end{center}
\footnotemark[*] The velocity difference derived from the XIS spectra.
The $68$\% statistical errors and (the 68\% systematic errors) are
quoted.  \footnotemark[$\dagger$] The velocity difference and the
$1\sigma$ error derived from the Chandra ACIS spectra
\citep{Dupke_Bregman_2006}.
\end{table*}
%%%%%%%%%%%%%%%%%%%%%%%%%%%%%%%%%%%%%%%%%%%%

As to (1), the model fitting to the XIS spectra has given
$z_{-}=0.0101\pm0.0005\,(\pm 0.0015)$, and
$z_{+}=0.0079\pm0.0012\,(\pm 0.0015)$, resulting in a negative value
for the redshift difference $\Delta z \equiv z_+ - z_- = -0.0022 \pm
0.0013\,(\pm 0.0022)$, or the velocity difference, $\Delta v = c\Delta
z= -660 \pm 390 \,(\pm 660)~{\rm km\,s^{-1}}$.  Because PEAK- and
PEAK+ are close to each other, the intrachip gain difference between
them is considered to be smaller than our canonical value, $\pm
470~{\rm km\,s^{-1}} $.  Nevertheless, to be conservative, we
multiplied this by $\sqrt{2}$, to obtain $\pm 660~{\rm km\,s^{-1}}$ as
our systematic error on the velocity difference.  The above results
then imply that the velocity difference between PEAK- and PEAK+ does
not significantly exceed the calibration uncertainty of $\pm 660~{\rm
  km\,s^{-1}}$.   If we further sum the systematic and statistical
  errors in quadrature, the 90\% confidence range becomes $-1890~{\rm
    km\,s^{-1}} < \Delta v < +570~{\rm km\,s^{-1}}$, which is
  inconsistent with the Chandra measurement of $2900\pm 700~{\rm
    km\,s^{-1}}$.  The results are summarized in table~\ref{tab2}.

As to (2), a similar analysis of the Suzaku XIS data has given
  the redshifts of the R-2 and R+3 regions as $z_{\rm
    R-2}=0.0107\pm0.0006\,(\pm 0.0015)$ and $z_{\rm
    R+3}=0.0089\pm0.0010\,(\pm 0.0015)$, respectively.  Therefore, the
  redshift and velocity differences become $\Delta z = z_{\rm
    R+3}-z_{\rm R-2} = -0.0018\pm 0.0012\,(\pm 0.0022)$ and $\Delta v=
  -540 \pm 360 \,(\pm 660)~{\rm km\,s^{-1}}$, respectively
  (table~\ref{tab2}).  The allowed 90\% range then becomes $-1730~{\rm
    km\,s^{-1}} < \Delta v < +650~{\rm km\,s^{-1}}$, which again
  excludes the Chandra measurement of $\Delta v$ between R-2 and R+3.

Then, can we reconcile the apparently conflicting Chandra (plus
  ASCA) and Suzaku results?  One possibility is that the much wider
  PSF of the Suzaku XRT, compared to that of Chandra, has diluted the
  velocity gradient.  To examine the effect, we again executed the
  {\tt xissim} simulation of the XIS spectra for PEAK- and PEAK+,
  simply assuming that the gas bulk motion exists only in those two
  regions and that they have the velocity difference close to the 90\%
  lower limit of the Chandra result, $\Delta v=+1700~{\rm
    km\,s^{-1}}$; other regions in the XIS field of view are assumed
  to have no bulk motion, with $z=0.0104$.  The model fit to the two
  simulated spectra indicates that the Suzaku XIS should observe a
  velocity difference of $\Delta v=+690~{\rm km\,s^{-1}}$, as a result
  of the telescope resolution which reduce the velocity difference by
  a factor of $\sim 2.5$.  Therefore, the Chandra measurement may be
  brought to be marginally consistent with the present Suzaku upper
  limit.  As to (2), the same simulation indicates that the 90\%
  lower-limit value of $\Delta v=+760~{\rm km\,s^{-1}}$ between R-2
  and R+3 based on the Chandra report will be reduced to $\Delta v
  =+300~{\rm km\,s^{-1}}$ in the XIS data after smeared by the PSF.
  This is consistent with the 90\% upper limit set by Suzaku,
  $+650~{\rm km\,s^{-1}}$.

In this way, the positive $\Delta v$ reported by
\citet{Dupke_Bregman_2006} for the two pairs of regions can be
tentatively made consistent with the negative $\Delta v$ we measured.
However, this was possible by invoking rather extreme assumptions;
namely, to employ the Chandra lower limit and the Suzaku upper limit
on $\Delta v$, and to assume that the ICM bulk motion is localized to
the small regions of $\sim 25$~kpc.

A comparison of the present results with the reported ASCA measurement
\citep{Dupke_Bregman_2001} may give additional clues.  The Suzaku XRT
\citep{Serlemitsos_etal_2006} is an improved version over its
predecessor onboard ASCA.  Similarly, the Suzaku XIS is an improved
instrument based on the SIS experiment onboard ASCA.  As compared with
the XRT+SIS combination onboard ASCA, the XRT+XIS of Suzaku hence has
a $\sim1.5$ times better PSF, a $\sim4.5$ times larger effective area
(at 7 keV; two SIS sensors versus three XIS cameras).  Furthermore,
the energy scale of the Suzaku XIS has been better calibrated
\citep{Koyama_etal_2006} than the ASCA SIS which was the first
single-photon detection CCD used in cosmic X-ray observations.
Therefore, any Doppler shifts in the Fe-K line detectable by ASCA
would be detected by Suzaku with a much higher significance (unless
the exposure is too short).

From these considerations, we conclude that it is rather difficult to
reconcile the the present results with the Chandra (plus ASCA)
measurements.  Additional information will be obtained by analyzing a
longer Chandra dataset on the Centaurus cluster.
\citep{Fabian_etal_2005}.
\footnote{Some of the present authors actually analyzed 
this data set, but did not detect any
significant velocity differences between the R-2 and R+3 
regions identified by \citet{Dupke_Bregman_2006}.}

\subsection{Implications of the derived upper limits}
\label{subsec:implications}

Utilizing the Fe-K$\alpha$ line, we obtained a 90\% upper limit on the
velocity difference of $|\Delta v| < 1400~{\rm km\,s^{-1}}$ between
any pair of regions within the XIS field of view, separated typically
by $\sim 50$~kpc or more.  This gives a tighter constraint on the ICM
bulk motion than was suggested by previous reports.  Since the ion
sound velocity in the ICM is $s\equiv(\gamma P/\rho)^{1/2}=(\gamma
kT/\mu m_p)^{1/2} \sim 880 ~{\rm km\,s^{-1}}$ for a temperature of
$kT=3$~keV, our results imply that the line-of-sight velocity of the
ICM bulk motion, if any, does not largely exceed the sound velocity.
In other words, the ram pressure associated with the ICM bulk motion
is at most comparable to the thermal ICM pressure, thus approximately
validating the assumption of hydrostatic equilibrium in calculating
the cluster mass.  Of course, we can probe the gas motion only along
the line of sight, and hence we need to assume 3-dimensional velocity
structures in order to quantitatively utilize the present upper
limits.

As an elementary exercise, below we evaluate the upper limit on
rotational motion assuming for simplicity that the gas is rigidly
rotating at a typical circular velocity of $\sigma_{r} \sim |\Delta
v|/2$, and derive its quantitative effect on cluster mass estimation.
Assuming an approximate spherical symmetry, and including the
centrifugal force, the balance against the gravitational pull at a
radius $r$ on the rotational equatorial plane then becomes
\begin{equation}
  -\frac{GM(r)\rho_{\rm gas}}{r^2}  
  = \frac{\partial P_{\rm gas} }{\partial r} 
  - \frac{f\rho_{\rm gas}\sigma_r^2}{r}, 
\label{eq1}
\end{equation}
where $M(r)$ is the cluster mass interior to radius $r$, $\rho_{\rm
  gas}$ is the gas density, $P_{\rm gas}=\rho_{\rm gas} kT (\mu
m_p)^{-1}$ is the ICM pressure, and $f$ is the fraction of the ICM
that is rotating ($0\leq f \leq1$).  The gas density profile of the
Centaurus cluster roughly follows $\rho_{\rm gas}\propto r^{-1}$
within the central $r\lesssim 10\arcmin$ region, assuming the
double-$\beta$ model \citep{Ikebe_etal_1999, Mohr_etal_1999}.
Together with the assumption $\sigma_r \propto r$, this yields
\begin{equation}
 f\rho_{\rm gas}\sigma_r^2 \propto r, 
\end{equation}
and hence the second term on the right-hand side of equation~\ref{eq1}
can be rewritten as
\begin{equation}
 \frac{f\rho_{\rm gas}\sigma_r^2}{r} 
  = \frac{\partial (f \rho_{\rm gas}\sigma_r^2)}{\partial r}.
\end{equation}
Equation~\ref{eq1} then becomes
\begin{eqnarray}
  -\frac{GM(r)}{r^2} 
  & = & \frac{1}{\rho_{\rm gas}} \frac{\partial P_{\rm gas} }{\partial r} 
  + \frac{1}{\rho_{\rm gas}} 
  \frac{\partial (f \rho_{\rm gas}\sigma_r^2)}{\partial r}\nonumber \\
  & =& \frac{1}{\rho_{\rm gas}} 
  \frac{\partial}{\partial r}P_{\rm gas} (1+f \beta_r), 
\end{eqnarray}
where 
\begin{equation}
  \beta_r \equiv \frac{\mu m_p \sigma_r^2}{ kT} 
  \sim 1.07 \left(\frac{\mu}{0.63}\right)
  \left(\frac{\sigma_r}{700~{\rm km\,s^{-1}}}\right)^2 
  \left(\frac{kT}{3~{\rm keV}}\right)^{-1}
  \label{eq4}
\end{equation}  
is the ratio of the kinetic energy due to rotational motion to the
thermal energy.  Thus, the total mass should be higher than the
hydrostatic mass by a factor of $(1+ f\beta_r)$, because of the
presence of additional pressure support.

At a typical radius of $\sim 100$~kpc, the rotational velocity is
$\sigma_r < |\Delta v|/2 = 700~{\rm km\,s^{-1}}$, yielding $(1 +
f\beta_r)\lesssim 2$.  Considering an inclination angle $i$ of the
rotation axis and the mean $\sin{i}$ factor, $2/\pi$, the rotational
velocity is $\sigma_r < |\Delta v|/2\sin{i} = 1100~{\rm km\,s^{-1}}$
and $(1 + f\beta_r)\lesssim 3.6$.  Therefore cluster mass estimation
with the assumption of hydrostatic equilibrium is justified within a
factor of $\sim 2-3.6$.  In order to put a stronger constraint on the
cluster mass to compare with other techniques (particularly
gravitational lensing), a two-fold improvement on the energy-scale
calibration is necessary.

\subsection{Turbulent broadening of iron lines}\label{subsec:turbulent}

The velocity structure in the cluster can also be studied by analyzing
the X-ray line widths for turbulent Doppler broadening of the X-ray
emission lines.  The iron ions provide the best tracers of this
effect, because the broadening due to a turbulent velocity, $\sigma
_{\rm turb}$, on the order of $\sim 300~{\rm km\,s^{-1}}$, can exceed
their thermal broadening ($\sim 2$~eV) by several times
\citep{Inogamov_Sunyaev_2003}.  However, considering the energy
resolution of the FI CCDs, 160~eV (FWHM) or 70~eV (Gaussian
$1\sigma$), what can be done with the present data is at most to
examine the observed Fe-K line width for possible broadening beyond
the instrumental energy resolution and its intrinsic line width
($\sigma_0\sim 28$~eV for $kT=3$~keV).  In order to carry out this
search under the highest statistics, we again fitted the
XIS0+XIS2+XIS3 spectrum from the central $18\arcmin\times18\arcmin$
region (analyzed in \S\ref{subsec:analysis_large}) with the APEC plus
three Gaussian model.  Here we assumed that the gas has no bulk
velocity, and fixed the centroid of the He-like K$\alpha$ line to
$6.677(1+z_{\rm cl})^{-1}$~keV.  As a result, the Gaussian width has
been obtained as $\sigma = 47\pm2$~eV.

Since the degradation of the energy resolution with time is not
currently implemented in the XIS response matrix
\citep{Koyama_etal_2006}, the fitted Gaussian width is expressed by
$\sigma^2 = \sigma_{\rm inst}^2 + \sigma_{0}^2 + \sigma_{\rm turb}^2$,
where $\sigma_{\rm inst}$ is the width of the Mn-K$\alpha$ calibration
line as mentioned before, and the average of the three FI CCDs gives
$\sigma_{\rm inst}=40\pm4$~eV.  Thus, the measured width is consistent
with that expected from the instrumental and intrinsic widths,
$(\sigma_{\rm inst}^2 + \sigma_{\rm 0}^2)^{1/2}=49\pm2$~eV,
  where the intrinsic width $\sigma_{\rm 0}$ includes also the thermal
  Doppler although it is negligibly small ($\sim 2-3$~eV) in the
  energy range of interest.  The turbulent broadening is then loosely
constrained as $\sigma_{\rm turb}< 32~{\rm eV}\sim 1400~{\rm
  km\,s^{-1}}$ (90\% confidence).  Although $\sigma_{\rm 0}$
significantly increases below a temperature of 2~keV, it is within the
range of $26~{\rm eV} < \sigma_0< 30~{\rm eV}$ as long as $2~{\rm
  keV}<kT<4~{\rm keV}$, and does not significantly affect the above
discussion.  The derived limit on $ \sigma_{\rm turb}$ is consistent
with a stronger constraint, $< 800~{\rm km\,s^{-1}}$, derived by an
argument that strong turbulence would smear out, via diffusion, the
observed abundance gradient in the central 100~kpc of the cluster
\citep{Graham_etal_2006}.

Observations of resonance scattering of the X-ray emission lines will
provide another way to constrain the ICM turbulent motion.  This is,
however, beyond the scope of the present paper.

\section{Summary}
Based on Suzaku XIS observations of the Centaurus cluster, we
investigated the bulk motion in the intracluster medium to constrain
its dynamical state.  The high sensitivity and accurate calibration of
the XIS sensors enabled us to study positional dependence of the
iron-K line energy on a 50~kpc scale, particularly over the central
$240~{\rm kpc}$ region, as well as on a $140$~kpc scale over a
diameter of 460~kpc.  The results show that there is no significant
velocity gradient, within the present calibration uncertainty of $\pm
660~{\rm km\,s^{-1}}$, suggesting that the bulk velocity does not
largely exceed the thermal velocity of the gas.  The 90\% upper
  limit on the velocity separation between a pair of regions is
  $|\Delta v|<1400~{\rm km\,s^{-1}}$, providing a tighter constraint
  than the previous observations by a factor of $\sim 2$.  The present
  result cannot be easily reconciled with the previous Chandra
  observations which reported the large velocity gradients in the two
  pairs of rectangular regions near the cluster core.  The mean
redshift of the ICM is determined to be $0.0097$, in agreement with
the optical redshift of the cluster within the calibrational error.
With the upper limit on the velocity difference, the pressure support
due to hypothetical rotational motion may not alter the cluster mass
estimate under the hydrostatic assumption by more than a factor of
$\sim 3$.

\bigskip
We are grateful to all the members of the Suzaku Science Working
Group, especially T. Tsuru and H. Nakajima for comments on the XIS
energy scale, Y. Ishisaki for useful suggestions on the ray-tracing
simulations, and N. Y. Yamasaki for discussions. N. O. acknowledges
support from the Special Postdoctoral Researchers Program of RIKEN.

\end{document}